# Molecular static simulation of edge dislocation core in bcc iron

A.A. Gusev[1,2], A.V. Nazarov[1,2, a]

[1] Institute for Theoretical and Experimental Physics named by A.I. Alikhanov of NRC "Kurchatov Institute", 25 Bolshaya Cheremushkinskaya str., 117218, Moscow, Russia

[2] National Research Nuclear University MEPhI (Moscow Engineering Physics Institute) 31 Kashirskoe shosse, 115409, Moscow, Russia

[a] avn46@mail.ru

**ABSTRACT**

We simulate the dislocation core structure in bcc iron using the modified Molecular Static method. A feature of this method is the application of an iterative procedure in which the atomic structure in the vicinity of the defect and the constants that determine the displacements of atoms immersed in the elastic continuum are calculated in a self-consistent manner. Following the mentioned approach, we develop a model for calculating the atomic structure of edge dislocations, taking into account the anisotropy of the elastic medium surrounding the main calculation cell. Anisotropy is taken into account by introducing an explicit angular dependence for the parameters of the elastic field created by the dislocation: magnitude of Burgers vector and Poisson's ratio. Simulation is carried out for a split dislocation with Burgers vector along [100]. The convergence of the iterative algorithm is shown and the influence of the computational cell size on the results is considered. Calculated results are: atomic structure of dislocation in bcc iron, angular dependence of the parameters describing the elastic dislocation field at large distances from the dislocation line, and the strain tensor components in the entire simulation area.

Keywords: edge dislocation, iron, simulation, modeling, molecular statics.

## 1. Introduction

Segregation of certain elements in the vicinity of the dislocation lines alters the structure of the defect core that affects the ductility of the material. Therefore, the study of the distribution of elements near linear defects is of interest. A necessary prerequisite for this is information on the atomic structure of the dislocation core. In this regard, we simulate the structure of the core of the edge dislocation, and its surroundings using the modified method of molecular statics, taking into account the anisotropy of the elastic field at the boundary of the computational cell. The developed model makes it possible to match the displacements of atoms at the boundary of the main computational cell with the displacements of atoms located in an elastic medium, and calculate the components of the strain tensor in the entire computational domain. The obtained strains are supposed to be used in the future for modeling the kinetics of the redistribution of impurity atoms in the dislocation field.

## 2. Model description

As a modeling method, the method of Molecular Statics is used [1], as in a number of other works [2-4]. In this case, the main computational cell in which the atoms relax is surrounded by an elastic medium (zone), the atoms in which are displaced relative to the nodes of the ideal lattice in accordance with the solutions of the equations of elasticity theory for such a defect as a dislocation [2]. As a rule, in the aforementioned works it is not possible to ensure the correspondence of atomic displacements at the boundary of two zones — the main computational cell and the elastic zone [4]. As a result of calculations, the difference in atomic displacements at the boundary of two zones can distort both the structure of the nucleus and the entire strain field in the vicinity of the dislocation.

Therefore, the applied model has significant differences from the previously used ones and is a development of the model that was previously applied to model point defects [5–8] and nanopores [9,10]. A common feature of these models is the application of an iterative self-consistent procedure for calculating the coordinates of atoms in the main cell and the parameters that determine the solutions of the equations of elasticity theory, the displacement of atoms immersed in an elastic medium. Moreover, the model used solutions for the case of dislocations in the approximation of an isotropic medium [2], and in the plane normal to the dislocation line we have:

$$u_x = \frac{b}{2\pi}\left(\operatorname{atan}\left(\frac{y}{x}\right) + \frac{\pi}{2}\operatorname{sign}(x) + \frac{xy}{2(1-\mu)(x^2+y^2)}\right), \quad (1a)$$

$$u_y = -\frac{b}{4\pi}\left(\frac{1-2\mu}{2-2\mu}\ln(x^2+y^2) - \frac{y^2}{(1-\mu)(x^2+y^2)}\right), \quad (1b)$$

where $b$ is Burgers vector, $\mu$ is Poisson ratio.

However, the crystal anisotropy is taken into account by determining the angular dependence of the Poisson coefficient and the Burgers vector determined during the iterations. Thus, the discrete nature of the crystal and the anisotropy of its properties are taken into account in the model. We add that the dislocation line is directed along the Z axis, and periodic boundary conditions are applied along this coordinate. System includes several atomic layers which are normal to Z axis. Burgers vector is oriented along X axis, and Y axis is oriented along line of extraplane's atoms. Simulation for every atomic layer is performed in similar way. The scheme for computational cell (one layer) and the main steps of the algorithm are shown in figure 1.

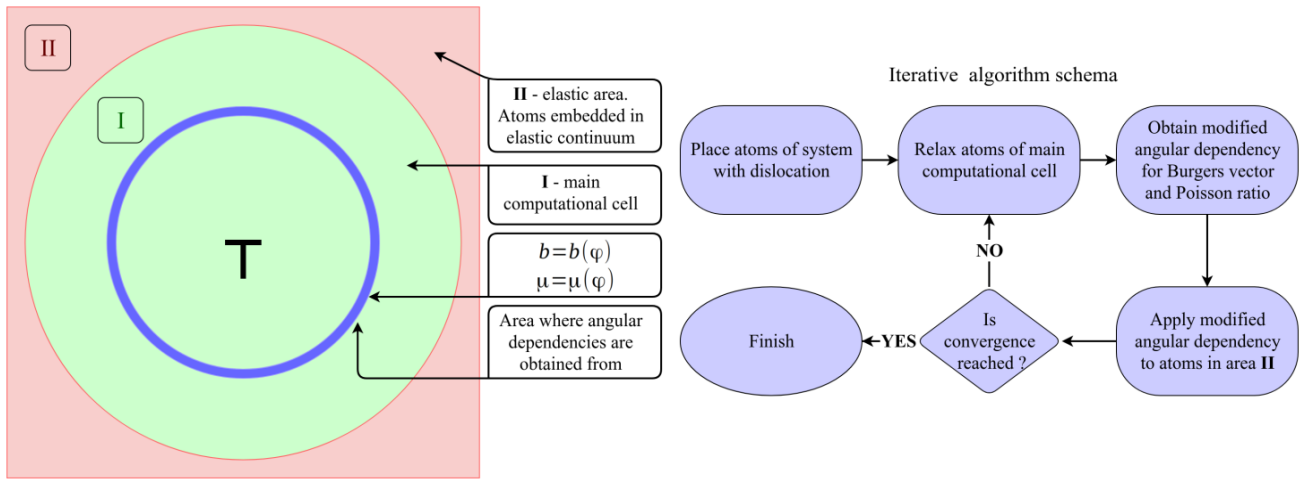

**Figure1.** The scheme of computational cell and the iterative algorithm steps

The equilibrium positions of the atoms of the main computational cell are obtained using the variational procedure, like in usual molecular statics method [1]. An important feature of the algorithm that distinguishes it from usual versions of the Molecular Static Method is the application of the iterative procedure (see Figure 1), in which the atomic coordinates of the main computational cell and two parameters are calculated in a self-consistent manner at each iteration step. These parameters are the Burgers vector and the Poisson ratio. Moreover, these parameters and their angular dependence are found on the basis of atomic displacements in the annular layer (see Figure 1) calculated at the next iteration step by solving the inverse system of equations 1a and 1b.

In this work, the structure of a split dislocation in bcc iron with the direction of the Burgers vector [100] is simulated using the interatomic potential of the EAM type [11]. A schematic structure of the dislocation is shown in Figure 2.

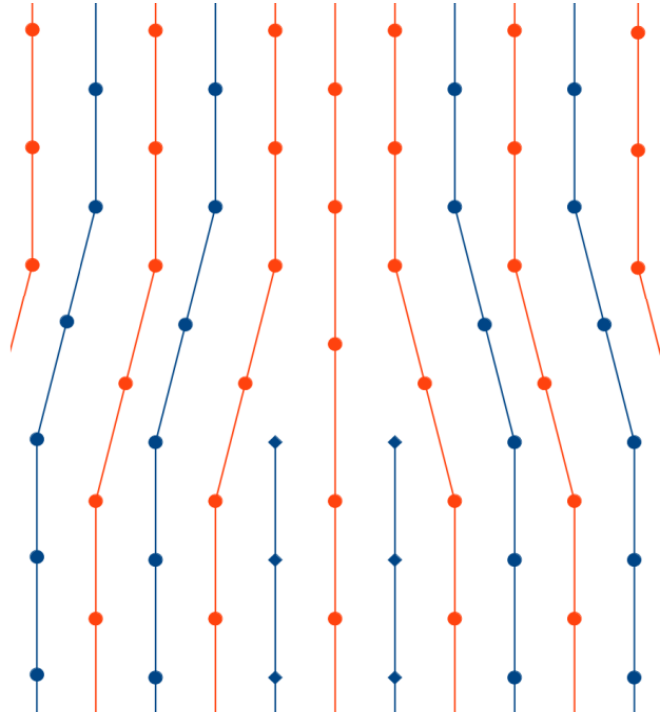

**Figure 2**. Scheme of a split dislocation structure

## 3. Results and discussion

We test the model by studying the dependence of the results on the size of the system and the number of iterations. Simulation is performed for systems of several sizes (106 Å x 106 Å, 123 Å x 123 Å, 264 Å x 264 Å). Figure 3 and figure 4 show the angular dependences of the Burgers vector and the Poisson ratio on the number of iterations. Zero angle is set for Y axis (corresponds with extraplane).

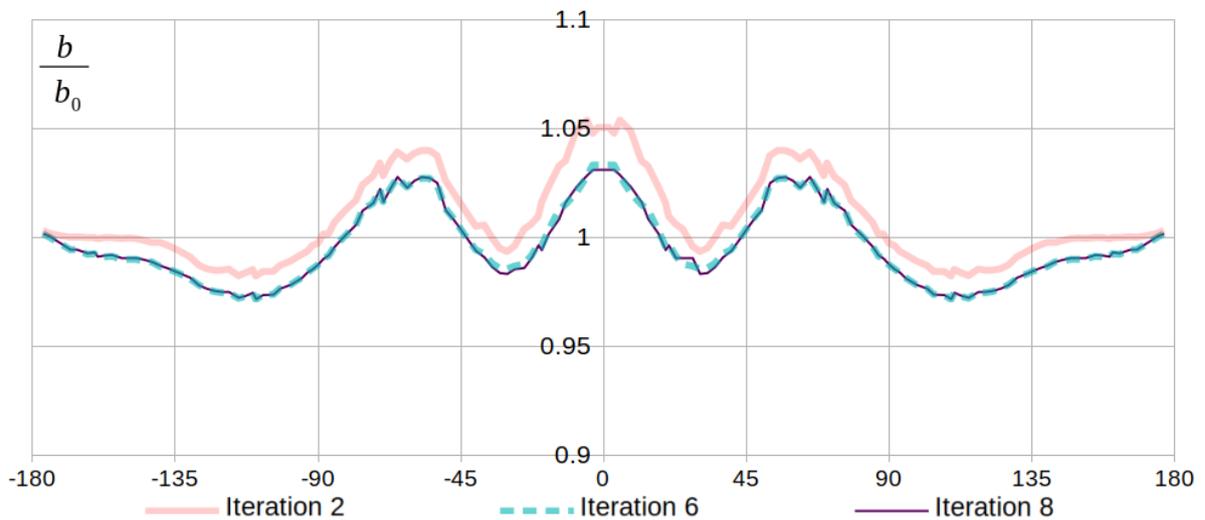

**Figure 3**. Normalized angular dependence of the Burgers vector magnitude for several iterations. $b_0 = 2.86$ Å

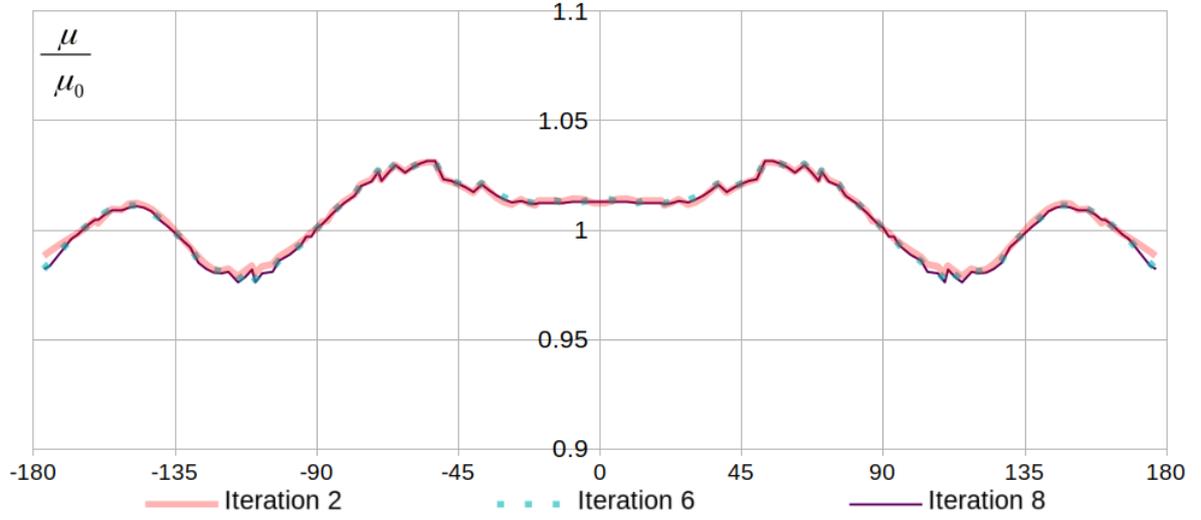

**Figure 4**. The normalized angular dependence of the Poisson's ratio for several iterations. $\mu_0 = 0.369$ [12]

Simulation shows that if size of the system is bigger than 132Å x 132 Å, then results are weakly dependent on size. The application of the developed model made it possible to calculate atomic displacements both in the main computational cell and in the elastic zone. It is worthy to emphasize that our approach grants good agreement of displacements at the zone boundary. Spatial distributions of the strain tensor components are calculated on the basis of displacements obtained during simulation, according to the known formulas:

$$\varepsilon_{ij} = \frac{1}{2}(\frac{\partial u_i}{\partial x_j} + \frac{\partial u_j}{\partial x_i}), \qquad (3)$$

where indices *i* and *j*, take values of 1 and 2 due to symmetry. The corresponding dependence $\varepsilon_{11}(x,y)$ and $\varepsilon_{22}(x,y)$ in a plane normal to a line dislocation are shown at figure 5 and figure 6.

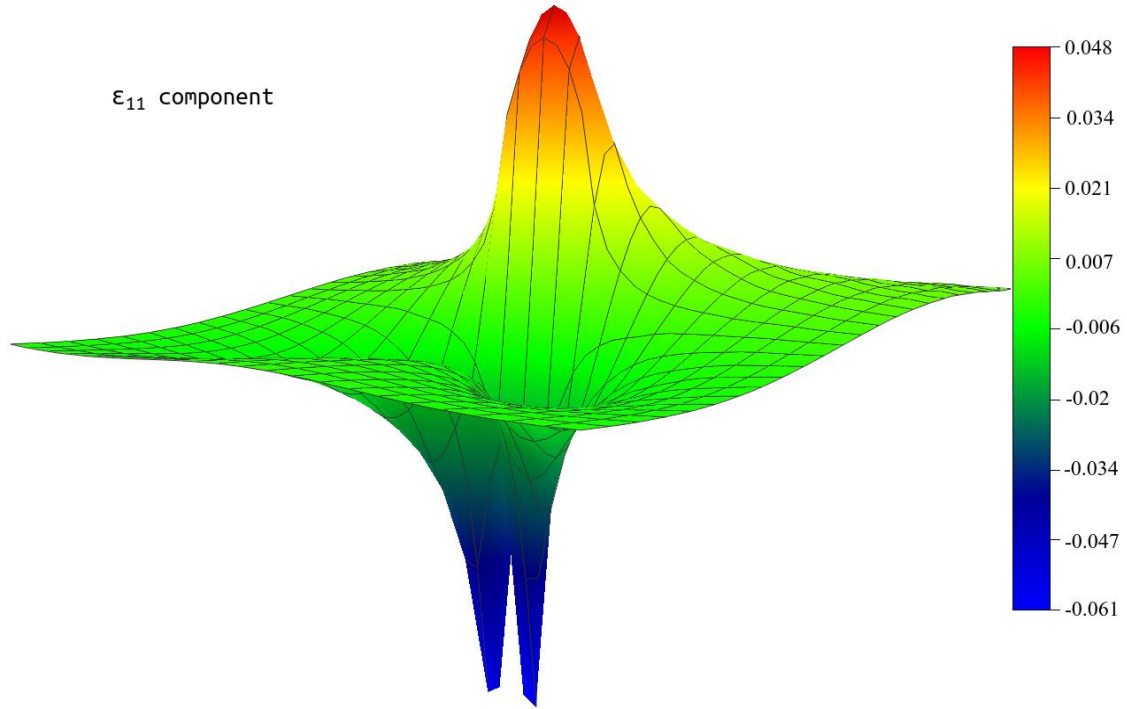

**Figure 5**. Spatial distribution of the tensor component $\varepsilon_{11}(x,y)$, *dl*=2.86

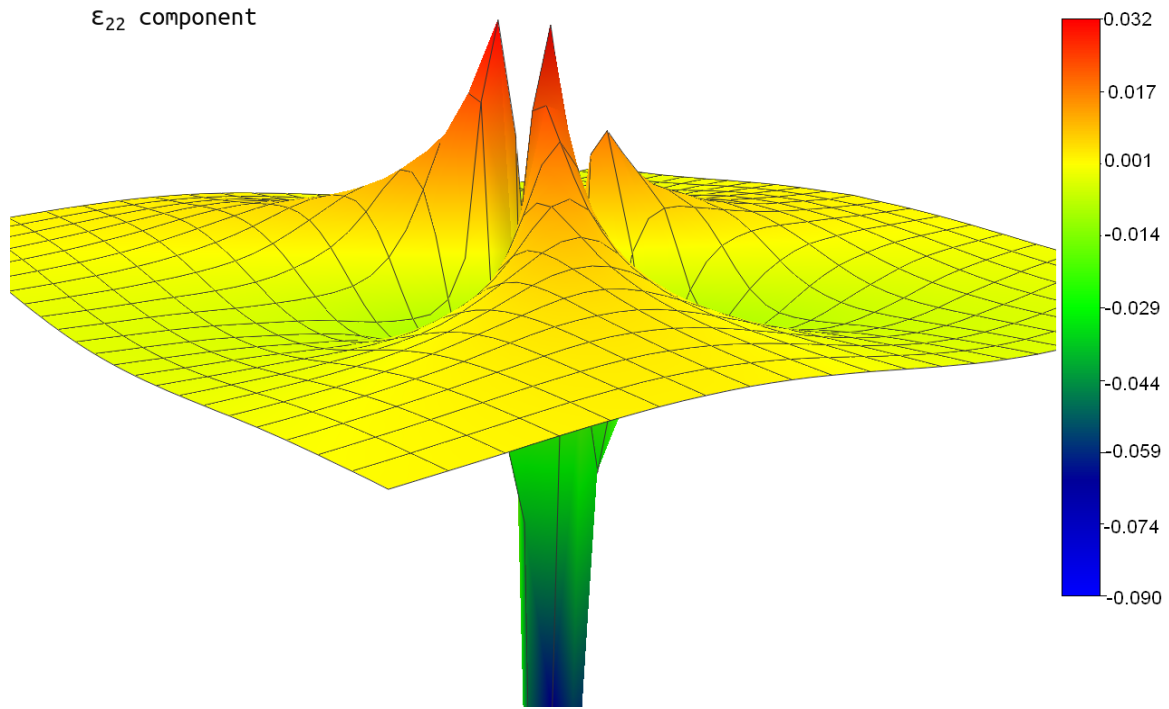

**Figure 6.** Spatial distribution of the tensor component $\varepsilon_{22}$ *(x,y)*, *dl* =2.86

Here *dl* is the step along the grid, the size of the shown area is 132 Å x 132 Å, and the calculations are performed for the area 264 Å x 264 Å. As it can be seen on figures, distributions have complex nonlinear character, especially near the dislocation line.

As it is indicated in [4], since the atoms embedded in elastic zone are usually fixed in their initial position and only the atoms of the main computational cell move, artifacts of elastic fields may take place at the boundary of these zones. The applied iterative algorithm modifies coordinates of atoms in zone II. This provides much better coordination of atomic displacements in the vicinity of the boundary between the main computational cell and the elastic zone. These features are very important in further solution of the equations that describe impurity atoms redistribution in the elastic field of dislocations, where diffusion coefficients are nonlinearly dependent on strain tensor components.

**4. Conclusion**
We developed model that is based on Modified Method of Molecular statics, that allows to calculate atomic structure of dislocation core and its vicinity. The feature of model is self-consistent iterative procedure for determining the coordinates of atoms in main computational cell and parameters that define displacements of atoms in elastic zone. Moreover, model allows taking anisotropy of the crystal structure into account, because these parameters depend on the angle.

The atomic structure of a split dislocation with a Burgers vector [100] in bcc iron is simulated, and strain tensor components are calculated.

The applied iterative algorithm gives good compatibility of atomic displacements in the vicinity of boundary between main computational cell and elastic zone.

Results of simulation show that distributions of strain tensor components in the vicinity of the dislocation line have a complicated nonmonotonic character.

**Acknowledgements**
Authors would like to acknowledge the financial support of the National Research Nuclear University MEPhI Academic Excellence Project (Contract No. 02.a03.21.0005).